# Effect of the repulsion between twin granular impactors on crater's aspect ratio: preliminary findings

P. Altshuler,[1] R. Pupo-Santos,[2] A. Rivera,[3] and E. Altshuler[1]

[1]*Center for Complex Systems, Physics Faculty, University of Havana, 10400 Havana, Cuba*
[2]*Espoleta Tecnologías, S. R. L., 32 No. 119, Miramar, 11300 Havana, Cuba.*
[3]*Zeolite Engineering Lab, IMRE, University of Havana, 10400 Havana, Cuba*

(Dated: December 21, 2025)

The formation and morphology of impact craters has been a subject of systematic attention by astronomers and geophysicists for a long time. Over the last years, *impact cratering* has become a notorious research subject for granular matter scientists, since it allows reproducing crater formation and intruder penetration processes by means of relatively inexpensive laboratory experiments [1–5].

Actual craters of geophysical and astronomical interest can be produced by the impact of a single body, but also by multiple impactors in parallel. The so-called "double-craters", for example, are believed to be caused by the impact of two solid bodies [6]. Those have been reproduced quite convincingly by table-top experiments involving spherical masses impacting a granular bed by Jiménez *et al.*, [7], Kitagwa *et al.* [8] and Yuxuan Luo *et al.* [9] –to the authors' knowledge. Interestingly, the latter incorporates the fact that two granular intruders interact repulsively with each other as they enter a granular medium [10–12].

Here we study the role of repulsive granular interactions in double crater formation from a different angle. We demonstrate experimentally that *repulsion does have a measurable effect in the shape of binary craters* under experimental conditions analogous to those reported in [7, 8]. Furthermore, we show that the protocol followed for the preparation of the granular bed plays a crucial role in the output of table-top experiments on doublet craters.

In our experiments, we dropped pairs of 11.0 mm diameter steel beads on the free surface of sand deposited into a cylindrical container (15.5 cm wide and 6.5 cm tall) from a height of 50 cm. The sand was collected at "Santa María" beach (Havana, Cuba); the grain size was distributed between 0.16 mm and 0.36 mm, peaking at 0.26 mm [13].

Two protocols were followed for the preparation of the sand bed before each dropping experiment. In the first one (POUR-LEVEL) the container was emptied and refilled by pouring sand from above, slightly surpassing the rim of the container, and then its surface was levelled by horizontally sweeping a plastic ruler. In the second one (POUR-SHAKE) the container was emptied and re-filled by pouring sand from above, and then it was laterally shaken for a few seconds until an even free granular surface was obtained. This implied a more compact granular bed than the one resulting from POUR-LEVEL, but we did not quantify the packing fractions.

Pairs of beads were magnetically released at the same time, side-by-side (i.e. the inter-center distance between the beads was one bead diameter). Two bead-pair arrangements were studied: the FREE-TWINS were released without any attachment between the beads

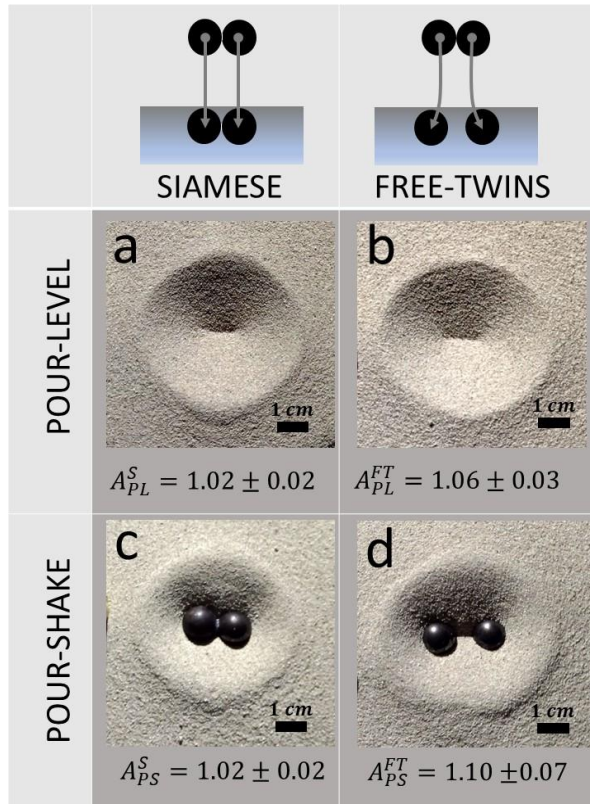

FIG. 1: Horizontal repulsion between twin intruders elongates granular craters. (a) Upper panel: image of a crater created by dropping two beads attached to each other from a heigh of 50 cm on a non-compacted sand bed. Lower panel: Aspect ratio calculated from 10 craters like the one shown in the image (b) Analogous data for craters made by non-attached beads on a non-compacted sand bed (c) Analogous data for craters made by attached beads, on a compacted sand bed (d) Analogous data for craters made by non-attached beds, on compacted sand.

$A_{PL}^{S} = 1.02 \pm 0.02$

$A_{PL}^{FT} = 1.06 \pm 0.03$

$A_{PS}^{S} = 1.02 \pm 0.02$

$A_{PS}^{FT} = 1.10 \pm 0.07$

(as in [7, 8]) while the SIAMESE were glued together by a small lateral area of the spheres. So, while FREE-TWINS could move horizontally relative to each other when touching the granular bed, SIAMESE could not. Ten droppings were performed for each combination of sand filling protocols and bead-pair arrangements, for a total of 40 experiments.

Fig. 1 comprises our basic results. Before discussing it, we should define the aspect ratio of the craters, as:

$$A = \frac{\lambda_\parallel}{\lambda_\perp} \quad (1)$$

where $\lambda_\parallel$ is the edge-to-edge length of the crater along a direction passing by the centers of the two impacting beads, and $\lambda_\perp$ is the edge-to-edge width of the crater perpendicular to the first direction [7]. Both were measured with a 0.05-mm resolution vernier caliper, while shining the craters with an inclined laser sheet in order to facilitate visualization of the craters' edges.

Fig. 1(a) shows the results for the SIAMESE beads dropped on the POUR-LEVEL bed of sand. The upper image shows a digitized picture of a typical crater, taken from top. The image suggests that the crater is quite circular: in fact, the corresponding aspect ratio averaged over 10 repetitions gives aspect ratio ~ 1.02.

The situation basically repeats for the case of SIAMESE dropped at a more compacted sand, as illustrated in Fig. 1(c), where $A^S_{PS} \sim 1.02$. So, when the intruders are released side-by-side *and are not allowed to move relative to each other*, the resulting crater is quite circular, which coincides with the findings in [7], but for non-attached intruders.

A surprising picture arises when we examine the case of non-attached intruders. Fig. 1(b) reports the results for FREE-TWINS dropped on loose sand resulting from the POUR-LEVEL protocol. The craters are clearly more elongated than for SIAMESE; the aspect ratio is ~ 1.06: as the beads touch the free granular bed, force chains in the region between the intruders are "charged", in such a way that they push the intruders away from each other [11, 12], resulting in a more elongated crater. The phenomenon is more clearly revealed for the FREE-TWINS and POUR-SHAKE combination, as illustrated in Fig. 1(d). Here, the more compacted sand bed implies stronger force chains and presumably a larger repulsion force in the horizontal direction, provoking a large gap between the beads (of the order of one bead diameter), which is clearly visible in the upper panel of Fig. 1(d). This results in the largest aspect ratio among the four kinds of experiments, of ~ 1.10. We speculate that this phenomenon has not been reported in previous works because of the use of relatively loose, lower-friction granular packings, where "strong" force chains are less relevant. However, more complex granular materials are needed to mimic actual scenarios [14].

In order to visualize the process of crater formation in the presence of repulsive forces between the intruders, we briefly discuss a video taken at 1000 fps (1920 pixels × 1080 pixels) of an experiment where FREE-TWINS are dropped on a sand bed prepared by the POUR-SHAKE protocol. The video was taken with a CHRONOS fast camera model CR-21-1.0-32C, using a LAOWA Lens FF 12 mm F2.8 D-Dreamer, at an angle of approximately 30 degress relative to vertical direction.

Fig. 2 (a) - (f) show snapshots taken from the video and 2 (g) contains a graph displaying the time dependence of the distance between the centers of the beads. In (a), the impactors have not touched the granular surface yet, and the distance between the center of the beads is 6 mm (i.e., one diameter), as shown in (g). In (b), the beads have started to touch the granular surface, and lateral repulsion has already showed up (the inter-center distance is now of the order of 14 mm, as shown in (g)); the outward motion of sand has started. The inter-center distance growth keeps increasing from (c) to (d), until the beads finally remain still and the crater is completely formed, as illustrated in (f). The maximum inter-center distance between the beads is slightly above 1 cm, meaning that the gap between the beads is substantial: approximately one bead diameter, as seen in (g).

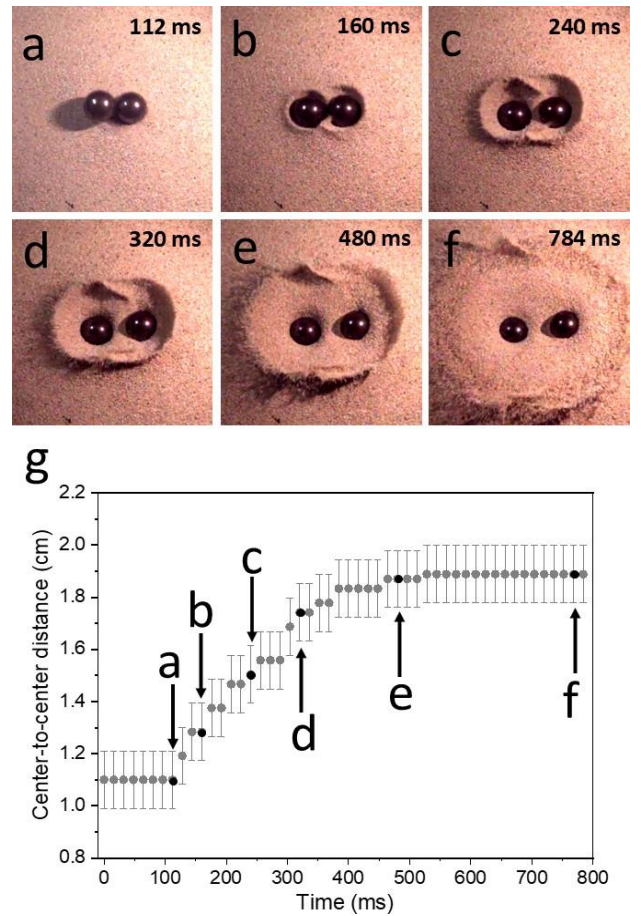

FIG. 2: Visualizing twin repulsion during impact cratering. (a)-(f) Snapshots for craters resulting from the impact of FREE-TWINS on POUR-SHAKE sand bed. (g) Evolution of center-to-center bead separation as a function of time

While Fig. 2 illustrates the role of repulsion for FREE-TWIN impacting POUR-SHAKE, repulsion is also visualized in analogous snapshots taken for FREE-TWIN hitting POUR-LEVEL sand (not shown here).

In conclusion, we have demonstrated that the repulsive horizontal interaction between granular intruders produces a measurable effect in the aspect ratio of the resulting crater, when they are near each other at the beginning of the impact event. This effect is more notorious when the granular bed is more compact.

### Acknowledgements

We acknowledge ARES project "*In situ* monitoring of water quality in Cuban bays: creating and promoting the use of a scientific toolbox" for material support. We thank R. Toussaint (ITES, University of Strasbourg) for introducing P. Altshuler to the use of *Blender* for image processing.